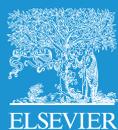

# Towards the development of human immune-system-on-a-chip platforms

Alessandro Polini[1,2], Loretta L. del Mercato[2], Adriano Barra[1,3,4], Yu Shrike Zhang[5], Franco Calabi[2] and Giuseppe Gigli[1,2]

[1] Dipartimento di Matematica e Fisica E. De Giorgi, University of Salento, Campus Ecotekne, via Monteroni, 73100, Lecce, Italy
[2] CNR NANOTEC – Institute of Nanotechnology c/o Campus Ecotekne, via Monteroni, 73100, Lecce, Italy
[3] INFN, Sezione di Lecce, Campus Ecotekne, via Monteroni, 73100, Lecce, Italy
[4] INdAM (GNFM), Sezione di Lecce, Campus Ecotekne, via Monteroni, 73100, Lecce, Italy
[5] Division of Engineering in Medicine, Department of Medicine, Brigham and Women's Hospital, Harvard Medical School, Cambridge, MA 02139, USA

Organ-on-a-chip (OoCs) platforms could revolutionize drug discovery and might ultimately become essential tools for precision therapy. Although many single-organ and interconnected systems have been described, the immune system has been comparatively neglected, despite its pervasive role in the body and the trend towards newer therapeutic products (i.e., complex biologics, nanoparticles, immune checkpoint inhibitors, and engineered T cells) that often cause, or are based on, immune reactions. In this review, we recapitulate some distinctive features of the immune system before reviewing microfluidic devices that mimic lymphoid organs or other organs and/or tissues with an integrated immune system component.

## Introduction

Tissue chips (TCs), OoCs, or microphysiological systems (MPSs) [1–3] are microfluidic *in vitro* systems that aim to replicate key structural and functional features of organs or tissue units in a convenient format for extended analysis and manipulation. Their objective is to provide standardized, economical, yet highly relevant *ex vivo* models to both enable the investigation of basic biological processes and improve the efficiency of the drug discovery process. By using patient-derived induced pluripotent stem cells (iPSCs) and their progeny as cell sources, they could ultimately lead to on-chip replication of an individual's disease for diagnostic and therapeutic testing and, hence, could become essential tools for personalized and/or precision therapies [4,5].

Numerous single-organ systems have been proposed, and efforts are under way to produce interconnected multiorgan platforms [6–8]. However, so far, the development of OoC systems to emulate the immune system has lagged behind, which is surprising for several reasons. First, immune mechanisms have a significant role in major diseases, including cancer, atherosclerosis, and the neurodegenerative diseases, in addition to being the primary cause of other common conditions, such as chronic infections and autoimmune diseases [9]. Second, modern approaches to molecularly targeted therapies often comprise complex entities such as monoclonal antibodies, stem cells, and nanoparticles, which are well above the size radar of the immune system. Such molecules can evoke unwanted immune responses, which can be catastrophic [10,11]. Third, selective activation or blockade of the immune system has emerged as a powerful therapeutic approach in its own right: significant results in at least some types of cancer have been reported for the use of immune checkpoint inhibitors [12], bispecific antibodies [13], and engineered T cells [14]. However, their use comes at a price, as exemplified by adverse effects such as the cytokine-release syndrome (CRS), and central nervous system (CNS)-related and immune-related adverse events (IRAE) [15]. The causes and remedies of these unintended reactions remain incompletely understood and represent an active area of investigation [16,17]. Thus, adequate models of the human immune system are needed to investigate its role in both pathogenesis and therapy.

Corresponding author: Polini, A. (alessandro.polini@nanotec.cnr.it)

Small rodents, such as mice and rats, have for decades provided convenient models for immunologists, and have been widely used for preclinical testing. However, they show significant differences with the human immune system [18]. In an effort to provide more human-relevant models, humanized mouse strains have been engineered that are engrafted with human hemopoietic precursors and populated mostly by mature human blood cells. These mice have proven useful in dissecting human-specific infections as well as IRAE [19]. However, they require special handling, are costly and pose ethical problems, quite apart from the fact that they do not yet fully reproduce the human system because of persistent mismatches [e.g., in major histocompatibility complex (MHC) antigens] between grafted human cells and host mouse cells. Thus, the development of a wholly human *ex vivo* immune system would provide significant added value.

In this review, we first briefly recapitulate some of the distinctive anatomo-functional features of the immune system and then report on the latest developments and applications of on-chip microfluidic devices specifically designed for the immune system.

## The human immune system: mechanisms and districts

The immune system has evolved to maintain the body homeostasis by recognizing and eliminating abnormal components whether of exogenous (microorganisms and macromolecules) or endogenous origin (e.g., cancer) [20]. Two interconnected subsystems may be distinguished: a phylogenetically older innate and a newer adaptive immune system. The former provides a fast response, but lacks fine specificity and memory, which are characteristics of the latter. Here, we describe some features and mechanisms of the human immune system: for a more comprehensive overview, please see Refs [9,20].

The innate immune system acts via preformed soluble proteins and phagocytic cells (macrophages and granulocytes) expressing germline-encoded receptors with broad specificity for shared molecular features of pathogens (pathogen-associated molecular patterns, PAMPs). Activated phagocytes not only kill microorganisms, but also release several factors (cytokines and chemokines) that trigger an acute inflammatory response and eventually the adaptive immune response. The latter proceeds via the generation of antigen-presenting cells (APCs, primarily macrophages and dendritic cells, DCs), their migration to local lymph nodes (LNs) and the priming of antigen-specific T and B cells. This leads to cell proliferation (clonal expansion) and differentiation and/or maturation into highly specific cellular effectors: from B cells into antibody-secreting plasma cells via steps that include isotype switching and affinity maturation; and from T cells into the various classes of helper ($T_H$), cytotoxic (CTLs), and regulatory ($T_{reg}$) T cells. T and B cell priming also gives rise to memory cells, which ensure a faster and more effective response upon re-encountering the same antigen.

In mammals, most cells of the immune system are produced in the bone marrow (BM) from multipotent hemopoietic stem cells (HSCs): most cells of the innate system derive from the myeloid lineage and most cells of the adaptive system from the lymphoid lineage. Unlike the former, most T and B cells undergo development and triggering in specialized structures. This is because their precursors clonally express large repertoires of random receptors deriving from somatic recombination and must undergo sequential steps of positive and negative selection to gate out cells with either excessive or insufficient affinity for self-molecules to ensure both self-tolerance and proper immune responsiveness. This mostly occurs in primary lymphoid organs (BM and thymus). By contrast, the secondary lymphoid organs (LN, spleen, and mucosa-associated lymphoid tissues, MALTs) are where mature T and B effectors of the adaptive immune response are generated.

Hemopoiesis depends on specialized BM niches that are organized into distinct functional microdomains [21,22]. Whereas early T cell progenitors migrate to the thymus, B cell development occurs entirely in the BM, and requires interactions with different types of resident stromal cell. Distinct niches in the BM promote homing of antigen-triggered B cells and the survival of antibody-producing plasma cells and of memory immune cells, while also providing a reservoir of myeloid-lineage immune cells. In the thymus, T cell development proceeds from the outer cortex to the medulla. Contact of T cell precursors with cortical epithelial cells has a key role in positive selection, and with APCs in the medulla in negative selection of self-reactive T cells.

The architecture of peripheral lymphoid organs is generally organized in distinct B cell and T cell zones, with specific classes of DCs. The former comprises follicles and germinal centers (GCs), which represent the areas where clonal expansion of antigen-triggered B cells occurs in parallel with affinity maturation (somatic hypermutation). LNs intercept the lymph draining from peripheral tissues and carrying tissue-derived antigen and APCs, while also receiving a constant supply of lymphocytes entering through the walls of specialized blood vessels, the high endothelial venules (HEVs), and leaving via the efferent lymphatics. Specialized cells (M cells) occur in the epithelia overlying MALTs and channel antigens from the lumen.

## Human immune-system-on-a-chip platforms: a long way to go

In recent years, the development of increasingly sophisticated and flexible microfabrication techniques has led to major progress in both basic and applied immunological research. This is particularly evident in biochip-based approaches to the manipulation of single immune cell populations for high-throughput analysis: currently, microfluidic devices are largely used to study the inflammatory responses *in vitro* down to the single cell level, providing interesting insights into the activation, adhesion, transmigration, phagocytosis, and secretion of immune cells (reviewed in Refs [23,24]). Here, we report the latest developments in those OoC platforms where an immune system component has been introduced and on progress relating to lymphoid OoC devices.

### Organ-on-a-chip platforms with an immune system component

Most of the work in this area has focused on three systems: (i) the interaction of immune cells with tumors; (ii) the interaction of leukocytes with endothelial cells; and (iii) inflammation.

Key advantages of tumor-on-chips with respect to other *in vitro* models, such as classical cultures and spheroids, are a controlled 3D architecture and dynamic microflow conditions, with exchange of oxygen and nutrients enabling operation for up to several days. In particular, reduced thickness and greater homogeneity with respect to spheroids facilitates both real-time imaging and drug and/or cell access to tumor cells [25]. The addition of immune cells, either as an adjacent extracellular matrix (ECM)-embedded compartment or perfused through a 'surrogate blood vessel', has allowed the detailed study by time-lapse microscopy of the different migratory phenotypes of macrophages and breast cancer cells [26] and of activated natural killer (NK) cells penetrating a glioblastoma tumor [27] in a manner that would have been impossible *in vivo*. Thus, the role of direct cell interactions versus paracrine signaling can be addressed. Moore and colleagues described a microfluidic model (*Ex Vivo* Immuno-oncology Dynamic ENvironment for Tumor biopsies; EVIDENT) for imaging the interactions between tumor fragments and flowing autologous tumor-infiltrating lymphocytes (TILs) over multiple days [28] (Fig. 1a). By automatic quantitative image analysis (Fig. 1b,c), the fraction of cell death attributable to TILs was estimated and found to respond to anti-PD-1 (an immune checkpoint inhibitor). Thus, the device has potential for the initial assessment of immunotherapeutics. Unlike animal models, on-chip systems are readily amenable to controlled manipulations of individual variables. The power and flexibility of the approach enabled Pavesi *et al*. to dissect the effects of low $O_2$ and inflammatory cytokines (IFN-γ and TNF-α) on the antitumor activity of engineered T cells, and to compare

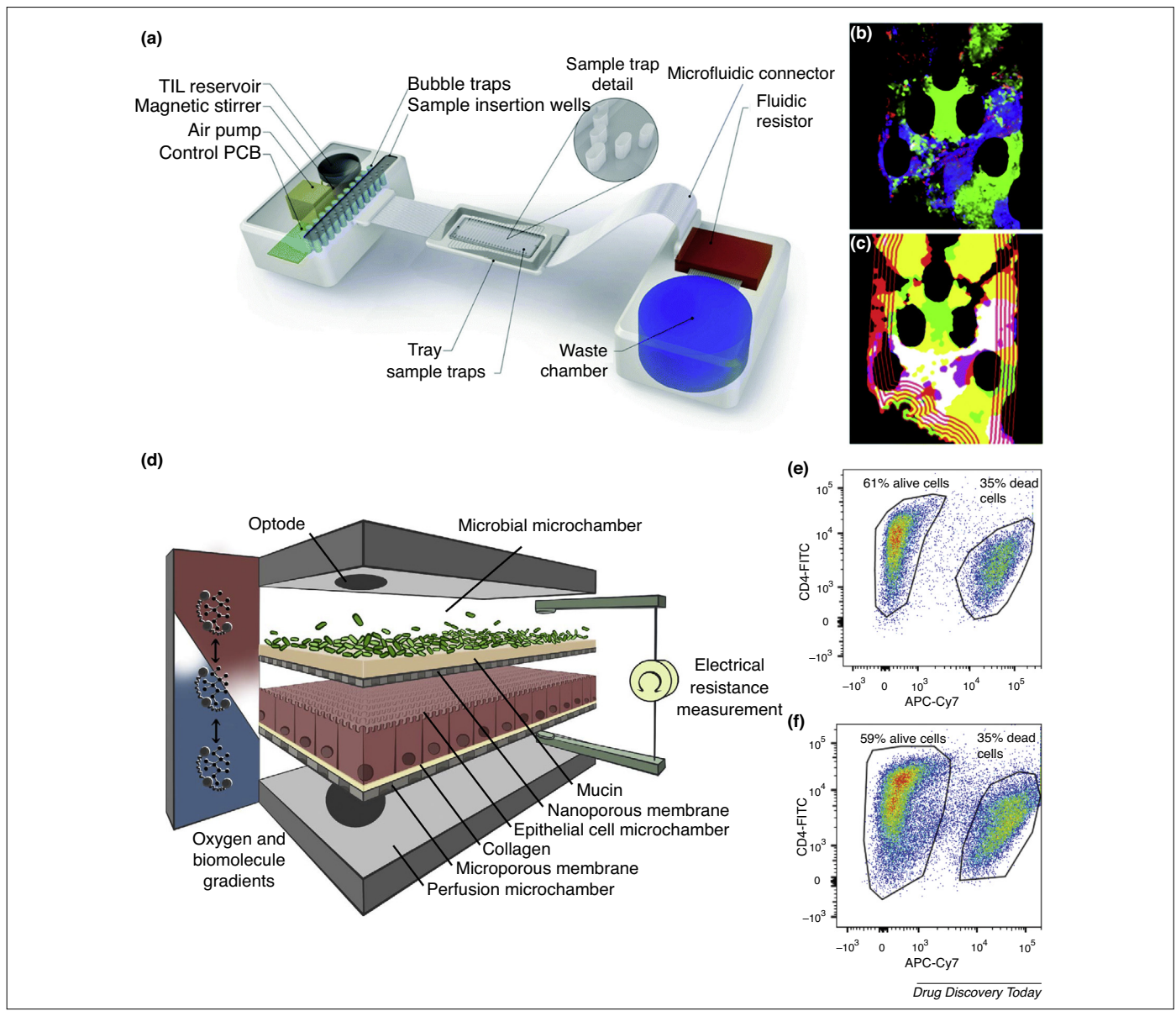

**FIGURE 1**

Examples of microfluidic devices for studying cell–cell interactions. (a) Illustration of the microfluidic model *Ex Vivo* Immuno-oncology Dynamic ENvironment for Tumor biopsies (EVIDENT) and its control system. (b) Map of tumor death through one plane of a z-stack confocal image. (c) Perimeter of tumor-infiltrating lymphocyte (TIL) infiltration versus time, showing the advancing front of TIL penetration into the tumor. (d) Conceptual diagram of the human-microbial crosstalk (HuMiX) device comprising three co-laminar microchannels: a medium perfusion microchamber; a human epithelial cell culture microchamber; and a microbial culture microchamber. Flow cytometry analysis of the viability of $CD4^+$ T cells cultured alone (e), or co-cultured for 24 h with LGG (*Lactobacillus rhamnosus GG*) (f). Reproduced, with permission, from Ref. [28] (a–c). Adapted, with permission, from Ref. [38] (d–f). Abbreviation: PCB, XXXX.

different technologies for T cell engineering [29]. Thus, screens can be devised to optimize the latter process. By contrast, by neglecting T cell migration, classical static 2D cultures provide misleading potency data.

Interactions between leukocytes and the endothelium have an important role in both the immune response and in triggering inflammation and vascular pathology (atherosclerosis). An inflammation-on-a chip device was realized comprising a central, endothelial-lined chamber flanked by ECM-filled compartments [30]. It was used to study the effects on neutrophil extravasation of precisely controlled and stable chemokine gradients and of variable ECM stiffness. Unlike classical Boyd chambers, the geometry avoids the confounding effect of gravity and provides high-resolution in situ imaging. A microvessel-on-chip has been described that reproduces lumen structure, barrier function, and secretion of angiogenic and inflammatory mediators, and can be maintained in culture for longer than 2 weeks [31]. Thus, it provides a highly relevant in vitro model, and was used to investigate transendothelial extravasation of both purified neutrophils and whole-blood cells in response to individual chemotactic factors. It could find application for the identification of both intrinsic and exogenous regulators of inflammation.

Rolling of activated T cells along the endothelial lining and their subsequent adhesion are affected by the local shear stress resulting from the blood flow rate and viscosity and from the vessel geometry. A dynamic microfluidic system to investigate the role of changes in the flow rate was reported by Park et al. [32]. A parallel multichannel architecture was designed to induce different cell adhesion molecules (CAMs) on human umbilical vein endothelial cells (HUVECs) and test the adhesion of primary T cells from blood. Adhesion of lymphocytes from patients with systemic lupus erythematosus (SLE) was higher than in controls, and was dramatically reduced by immunosuppressants. The device was proposed as a tool for high-content screening of novel drugs that requires less time, cost, and labor than conventional methods, while using human primary cells. Wu et al. realized a capillary-endothelium-mimetic microfluidic chip comprising a silicon micropore array lined on one side with an endothelial cell line and sandwiched between two microchannels [33]. It enabled the authors to investigate the effects of flow rate and chemotactic gradients on leukocyte extravasation over an extended period of time. Chemotactic factors were found to predominate at low flow rates, whereas extravasation was limited at higher flow rates because of cell aggregation near the side wall resulting from hydrodynamic forces. The device is also applicable to the study of cancer metastasis, atherosclerosis, and other angiopathies.

To simulate atherosclerosis, Menon et al. devised an original approach to pattern microchannels of complex geometry in micropillar-free ECM [34] and studied the effect of luminal stenosis on leukocyte and platelet interactions with normal and inflamed endothelium under different shear stress.

A humanized dynamic in vitro model based on polypropylene hollow microfibers featuring transmural microholes (2–4 μm) under pulsatile flow was designed to investigate the transendothelial trafficking of immune cells across the blood–brain barrier (BBB) [35]. In the presence of circulating monocytes, flow cessation followed by reperfusion caused BBB opening with abluminal extravasation of monocytes paralleled by a significant increase in proinflammatory cytokines and activated matrix metalloproteinases.

A microfluidic device (kit-on-a-lid-assay, KOALA) was described for the rapid purification of neutrophils from nanoliter volumes of blood and the reproducible generation of chemotactic gradients [36]. Its simple operation and versatility are useful for both clinical and research applications.

Other categories of microfluidic devices have been specifically developed for the study of inflammation. For example, interactions between the gut microbiome, intestinal mucosa, and immune components, and reduced peristalsis are believed to contribute to inflammatory bowel disease (IBD). Independent control of these factors in either animal studies or standard in vitro models would be difficult, if not impossible. Therefore, a physiologically relevant human gut-on-a-chip microfluidic device was developed that replicated the architectural and histotypical features of the human intestine in addition to peristalsis-like motions [37]. It enabled the co-culture of gut microbes in direct contact with the intestinal epithelium for more than 2 weeks, in contrast to static Transwell cultures or organoid cultures. Using this device, immune cells were found to synergize with lipopolysaccharide (LPS) or bacteria in inducing damage to both villi and the epithelial barrier through secretion of a set of proinflammatory cytokines. By manipulating individual factors (purified cytokines alone or in different combinations), a requirement for high levels of IL-8 for tissue damage was established. Moreover, bacterial overgrowth was found in the absence of cyclic mechanical deformations, even with constant luminal flow. This human gut on-a-chip could be further developed by incorporating primary or iPSC-derived epithelial cells, various subsets of immune cells, and microbial populations, and applied to identify and test potential therapeutic targets and drug candidates for IBD.

A different gut-on-chip was described by Shah et al. [38], named human-microbial crosstalk (HuMiX) and designed to allow aerobic conditions for human cells and anaerobic conditions for bacteria. As shown in Fig. 1d, it comprises three separate microchannels, one for perfusion, one for the human epithelial cells, and one for the microbes. Primary $CD4^+$ T cells were cultured for over 48 h in the presence of a facultative anaerobe commensal without any significant decrease in viability (Fig. 1e,f).

Inflammation is also at the root of common respiratory diseases, such as asthma, chronic bronchitis, and emphysema. The clinical relevance of animal models has been questioned because of structural and functional peculiarities, whereas standard in vitro 2D cultures generally lack properly arranged endothelial and immune system components and an active, shear stress-inducing fluid flow. To overcome these limitations, a human 'small airway-on-a-chip' was fabricated comprising an upper (airway) channel and a lower (microvessel) channel separated by a thin, porous polyester membrane colonized with lung airway epithelial cells and lung microvascular endothelial cells, respectively [39]. The device reproduced several structural and functional features of bronchioles for periods of weeks while enabling independent control of system parameters and analysis of human organ-level responses in real time with molecular-scale resolution. For example, both independent and collective responses of the lung epithelium and endothelium to a viral mimic [poly(I:C)] were identified, which would have been impossible in humans. The model was also used to investigate new therapies. Notably, the action of a novel inhibitor of nuclear factor (NF)-κB signaling in suppressing neutrophil adhesion was found to be largely microflow dependent, suggesting preferential inhibition of early events of the neutrophil-recruitment cascade, and consistent with a reduction in the expression of adhesion molecules on endothelial cells.

### Lymphoid organ-on-a-chip platforms

Although not ideal to study the complexity behind the lymphoid organ physiology and pathology, 2D protocols for static cell cultures have been known for a long time and have had a major role both in research and clinical settings: for example, long-term bone marrow cultures, LTBMCs, have been essential to study long-term *in vitro* myelopoiesis and lymphopoiesis [40,41]. More recently, 3D scaffold-based perfusion systems have been introduced for the study of BM [42,43]. Among the applications of microfluidic technology, an engineered hemopoietic bone marrow-on-a-chip (eBM) was produced by implanting into mice a cylindrical polydimethylsiloxane (PDMS) device containing a collagen scaffold enriched with bone morphogenetic proteins and demineralized bone powder, and with a single open end placed in the proximity of muscle (Fig. 2). This resulted in the formation of a shell of cortical bone surrounding hemopoietic marrow. Constructs transferred into a microfluidic device were maintained for up to 7 days on-chip with no decrease in hemopoietic stem/precursor cells (HSPCs) compared with fresh BM, as demonstrated by phenotyping and functional reconstitution of both myeloid and lymphoid lineages in γ-irradiated mice [44]. Moreover, the effect of γ-irradiation on the eBM was closely related to that on *in vivo* BM, unlike that on 2D cultures, proving the efficacy of this approach as a BM *in vitro* model.

Aiming at improving the lifetime of such artificial *in vitro* models for long-term testing, Sieber *et al.* achieved the successful long-term culture of HSPCs (up to 28 days, far longer than the conventional 7 days reached by other approaches) with full multi-lineage differentiation by means of co-cultivation with human mesenchymal stromal cells in a hydroxyapatite-coated zirconium oxide scaffold inside a multiorgan microfluidic platform [45].

Although no studies have focused on the adaptive immunity niches at the BM level, efforts have been devoted to modeling BM malignancies, such as leukemia and multiple myeloma, in microfluidic devices. For example, a 3D microfluidic acute lymphoblastic leukemia model, where leukemic cells were introduced along with stromal cells and osteoblasts in a collagen gel, showed how different drug responses are elicited by changes in the microenvironment [e.g., mechanical forces (flow) and biochemical cues] [46]. BM models can be also used as innovative tools for testing cellular immunotherapy approaches. Mesenchymal stromal cells and their osteogenic progeny were co-cultured with endothelial precursors in a 3D BM niche model facilitating the stable outgrowth of primary CD138$^+$ myeloma cells for up to 28 days [47]. The model was analyzed to assess the effects on primary myeloma cells of αβ T cells engineered to express a tumor-specific γδ T cell receptor (TCR).

In contrast to the BM, there are no published studies on engineering thymus-like architectures. The latest achievements in reproducing organ-like structures *in vitro* following cell biology approaches have developed, among many other organs, an organized and functional thymus-like organoid, able to support the development of CD4$^+$ and CD8$^+$ T cells: this was generated by reprogramming mouse embryonic fibroblasts into thymic epithelial cells *in vitro* with the upregulation of transcription factor forkhead box N1 (FOXN1) [48]. Further studies have to be designed to interrogate these thymus-like organoids to obtain fundamental knowledge on the function of the whole organ and, hopefully, to use these artificial structures in drug-testing platforms.

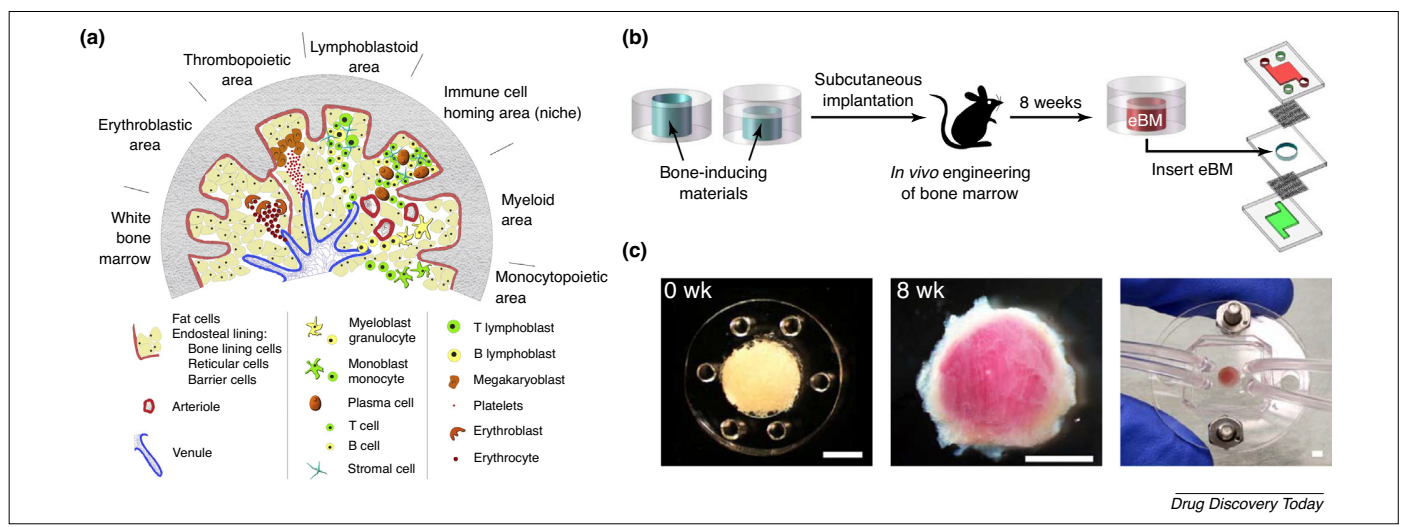

**FIGURE 2**

Bone marrow (BM) structure and an example of a BM-on-a-chip platform. (a) Schematic cross-section of the BM highlighting the different hematopoietic areas (where platelets, erythrocytes, lymphocytes, monocytes, and granulocytes are continuously created and released into the blood), the immunological niches (where memory T and B cells locate), and the large vasculature, necessary for the massive cell movement in the BM. (b) An engineered BM (eBM) was produced by implanting a polydimethylsiloxane (PDMS) device *in vivo* and later transferring it into a microfluidic platform. (c) A bone-inducing material is placed in a PDMS structure (left), implanted for 8 weeks to form a visible pink marrow (center) and then integrated in a microfluidic system (right). Scale bars: 2 mm. Adapted, with permission, from Ref. [75] (a) and Ref. [44] (b,c).

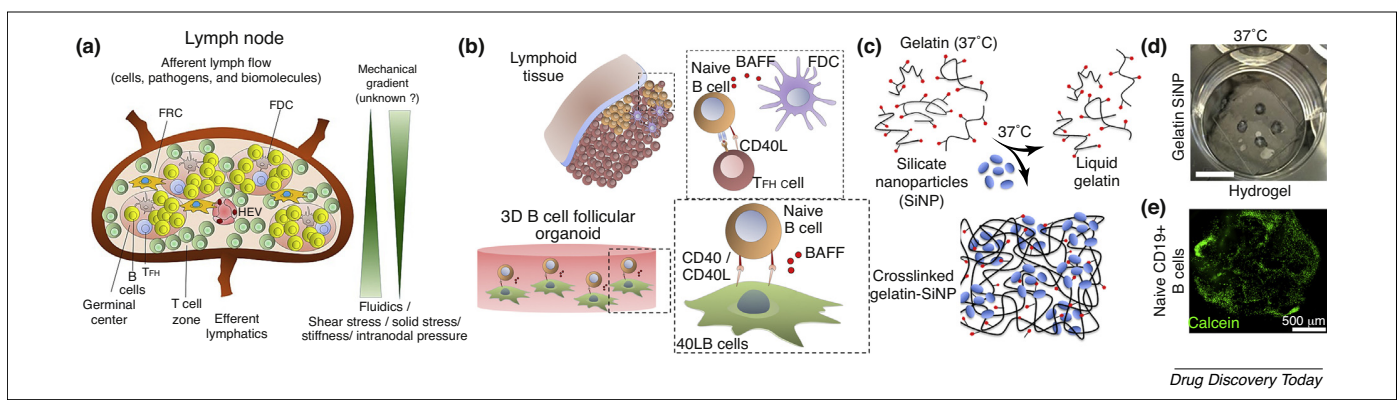

FIGURE 3

Lymph node structure and an example of an *ex vivo* engineered B cell follicle organoid. (a) Schematic of a lymph node showing the vessels that allow the in-and-out movement of lymphocytes, antigen-presenting cells, pathogens, and biomolecules. Specialized antigen-sampling, T cell, and B cell zones are indicated. (b) Schematic of the *in vivo* follicular interaction between mature naïve B cells, follicular T helper ($T_{FH}$) cells and follicular dendritic cells (FDCs), supporting the maturation of naïve B cells through the B cell activation factor (BAFF). (c) The cells were encapsulated in a silica nanoparticle-gelatin composite, which crosslinks at 37 °C and allows a proper B cell viability as shown in (d) and (e) (green calcein marking active B cells. Scale bar: 10 mm. Adapted, with permission, from Ref. [78] (a) and Ref. [50] (b–e). Abbreviations: FRC, fibroblastic reticular cells; HEV, high endothelial venules.

Most research on secondary lymphoid organ equivalents has been targeted towards LN-like architectures (Fig. 3a). The spatial complexity of the LN has been investigated in a microfluidic device capable of addressing individual B and T cell zones in an *ex vivo* slice of an isolated mouse LN [49], allowing local drug administration. Given the complexity of the LN architecture and functions, tissue-engineered LNs are currently designed to mimic only a few aspects rather than a complete artificial organ. For example, the possibility to replicate GCs *in vitro* provides an efficacious tool for understanding the maturation of B cells and designing innovative antibody-based therapeutics. In this context, Purwada *et al.* encapsulated murine naïve B cells and CD40L- and BAFF-expressing stromal cells into an arginylglycylaspartic acid (RGD)-functionalized gelatin hydrogel, reinforced with silicate nanoparticles for modulating the scaffold stiffness and porosity (Fig. 3b–e) [50,51]. The presence of the 3D structure potentiated CD40L/BAFF signaling and, upon addition of exogenous interleukin-4, led to a ~5–10-fold increase in the proliferation of $CD19^+$ $GL7^+$ GC B cells compared with conventional 2D co-culture systems, and a corresponding or higher increase in isotype-switched immunoglobulin (Ig)G1 cells by day 6. By replacing gelatin with a synthetic polymer displaying integrin ligands at a controlled density, the system was adapted to investigate the differential role of $\alpha_4\beta_1$- and $\alpha_v\beta_3$ integrins in the differentiation of GC B cells [52]. In a lymph node-on-a-chip flow device, a simple perfusion system applying a controlled tangential shear was introduced for the *in vitro* study of the mechanical forces between antigen-presenting DCs and different classes of T cells ($CD4^+$ versus $CD8^+$) or antigen-specific and unspecific T cells, undetectable with traditional *in vitro* systems [53]. Perhaps the most elaborate prototype of a human artificial lymph node (HuALN) was realized by a miniaturized, membrane-based perfusion bioreactor, hosting a hydrogel matrix preloaded with DCs through which T and B lymphocytes are made to recirculate continuously. A planar set of microporous hollow fibers provides continuous nutrient and gas exchange. Upon *in vitro* immunization, cell proliferation, antigen-dependent local cytokine release, assembly of lymphoid follicle-like structures (albeit not typical GCs), and an increase and decrease in IgM production mimicking a primary response were observed [54]. This model was further improved by the addition of mesenchymal stromal cells [55] and used to evaluate the efficacy of single vaccine candidates [56].

Efforts have also been made to realize cancer-on-a-chip models of solid lymphoid malignancies for drug discovery or mechanistic studies. A lymphoma-on-a-chip model was obtained by seeding tumor cells in a vascularized hyaluronic acid hydrogel and used to investigate the crosstalk between tumor cells, immune cells, and endothelial cells as well as the response to drug treatment [57]. A microreactor specifically engineered to provide uniform flow patterns and shear stress with a pressure gradient close to physiological values was used to show that fluid flow upregulates surface Ig and integrin receptors in subsets of diffuse large B cell lymphomas (DLBCL) via a mechanism mediated by CD79B and signaling targets [58].

The spleen is connected directly to the blood and has a central role in preventing sepsis conditions by cleansing the blood of pathogens, in maintaining erythrocyte homeostasis by removing older erythrocytes, as well as in the maturation of T and B lymphocytes. Whereas the first and second functions have been mimicked successfully in microfluidic devices, the third has not. A blood-cleansing microfluidic device has been designed for sepsis therapy [59]: it exploits the selectivity of magnetic nanobeads coated with an engineered human opsonin [mannose-binding lectin (MBL)] for removing a range of pathogens and toxins from the blood of an infected individual without activating complement factors. Its efficacy was demonstrated in a rat model of endotoxemic shock, where it increased animal survival rates after a 5-h treatment. Technological efforts have been also devoted to reproducing in a miniaturized system the peculiar hydrodynamic forces and physical constraints present in the spleen and fundamental for its filtering activity. Different devices [60,61] have been designed and successfully tested as *in vitro* platforms for assessing the deformability of red blood cells (RBCs) in different conditions, such as old and fresh RBCs, *Plasmodium*- parasitized RBCs, and RBCs from patients with hereditary spherocytosis.

Aiming at the creation of a versatile, multifunctional *in vitro* assay platform for studying the humane innate and adaptive immune responses within one system, the modular immune *in vitro* construct (MIMIC®) technology proposes a modular approach comprising three different modules [62]: (i) the peripheral tissue equivalent (PTE) module, resembling the innate immune response that occurs in peripheral tissues (e.g., skin and muscles), gives insights into the toxicity and immunostimulatory potential of a variety of biological and chemical compounds; (ii) the lymphoid tissue equivalent (LTE) module, which can include activated DCs from the PTE module, T cells, B cells, and follicular DCs, generates activated T cells, antibodies, and cytokines; and (iii) the functional assays or disease model module, which includes specific functional *in vitro* tests for studying vaccine candidates and pathogens.

## Theoretical perspectives and concluding remarks

An interesting application of immune-system-on-a-chip devices is the modeling of cell network properties. Thus, parameters extracted from extended imaging data sets of operating devices could be analyzed with mathematical tools from stochastic processes and statistical mechanics to provide a quantitative and predictive description of immune cell behavior. In one approach, time-lapse microscopy was used to examine the motility of either immunocompetent (wt) or immunodeficient [interferon-regulatory factor-8 (IRF-8)-knockout] splenocytes in the presence of melanoma cells in a two-chamber microfluidic device [63]. From images recorded at a sampling frequency of 1 every 4 min over 48 h and with a spatial resolution of 0.1 μm, several parameters were extracted: step length, time correlations, mean displacement, tortuosity, and ergodicity. The behavior of IRF-8-knockout splenocytes was characterized as a simple random walk, lacking any evidence of collective organization even in the presence of the target. By contrast, wild-type splenocytes could be modeled by biased random walks with a mean displacement of the ensemble growing linearly with time, which supports the hypothesis of a highly coordinated motion for the system as a whole. The analysis was extended to model the interaction between human peripheral blood mononuclear cells (PBMCs) and drug-treated breast cancer cells [64]: wild-type cells performed biased random walks towards the target, whereas heterozygous mutants heterozygous for formyl peptide receptor 1 (FPR1, a receptor for annexin A1 on dying cells) showed a weaker bias, and homozygous mutants performed uncorrelated random walks. Moreover, by performing stability analysis, novel reliable descriptors (Lyapunov coefficients) of both stable and metastable interactions between PBMCs and cancer cells were demonstrated. The motility analysis provided a quantitative correlate of the response to cytotoxic chemotherapy, which highlights the role of the immune system in this treatment. Thus, such an approach could be applied to predict the efficacy of chemotherapy in individual patients.

Microfluidic systems could also provide experimental tools to validate theoretical models whereby the balance between an effective immune response to foreign antigens and the suppression of autoimmunity depends on the degree of connectivity of the B cell network [65] or on collective decisions by the T cell population via a requirement for locally controlled quorum thresholds [66–69]. Furthermore, although necessarily replicating a limited fraction of the whole immune system, microfluidic devices could provide a sufficient set of averaged parameters to enable building of maximum entropy models predictive of the collective properties of the whole ensemble [70]. An intriguing possibility is to use such systems to create real, immune cell-based, pattern-diluted networks with experimentally tunable variables to enable different degrees of parallel processing [71,72]. The operational output of such 'immune processors' (i.e., either the relative expansion of individual clones or more sensitive indicators) might not only validate theoretical predictions, but also suggest novel approaches to manipulate the immune system.

Although the need to add immune system components to OoC platforms has been repeatedly underlined [7], achievements so far have been limited. This might be due in part to competition with humanized mice, which are likely seen by immunologists as a more relevant model, and partly to the complexity of the immune system itself, with its wide structural and functional diversity. Further progress calls for a closer collaboration between professional technologists and immunologists.

As reviewed here, a first generation of devices with limited functionality has been reported. Not surprisingly, they target preferentially the innate, fast response system (macrophage and neutrophils) rather than the slower and more complex adaptive system [73,74]. There is some progress in replicating myelo- and lymphopoiesis (BM) and antigen-priming sites in LN (with evidence of a primary immune response). However, no OoC platform exists that reproduces the key steps of repertoire selection as well the generation of functionally distinct sublineages (e.g., $T_H1/T_H2/T_H17$, $T_{FH}$ and $T_{reg}$).

Designing a functional immune system on microchips is clearly a daunting task of extreme complexity. The human immune system has been estimated to comprise $\sim 5 \times 10^{11}$ lymphocytes, of a similar order of magnitude to the number of neurons in the brain. Ig and TCR repertoires in mature lymphocytes are of the order of $10^8$, with the average size of a virgin (unprimed) clone being $\sim 10^2$–$10^3$ cells. Microfluidic systems generally handle sub-ml volumes, with fewer than $10^6$ cells. It has been argued that the limitation of cell numbers and, hence, of receptor repertoires, could strongly bias or skew the results [75]. This is clearly a key criticism that could be addressed by theoretical and experimental investigations of the (likely considerable) extent and functional consequences of redundancy in the immune system, as in other biological networks [76,77].

As briefly sketched above, the human adaptive immune system comprises a vast number of integrated components in a logistically highly distributed, yet precise spatial organization. The anatomical connections between individual parts, be they molecules or cells, are highly specific and must enable appropriate, efficient cell–cell interactions at both short range (e.g., during development and antigen presentation) and long range (the homing and recirculation of T cells). Although a microfluidic device could emulate a functional unit of the BM, thymus, or LN, dozens of devices are likely required to simulate even a rudimentary immune system. Moreover, they must be peripherally interfaced with all other organs, including the microbiota, given that fundamental inputs to the collective system are provided by the latter. However, although in-depth investigations of basic immunology are likely to require complex systems, basic systems, such as those illustrated here, offer potential over the shorter term, particularly for the initial assessment of novel personalized immunotherapies.

## Acknowledgments
This work was supported by the Progetto FISR – C.N.R. 'Tecnopolo di Nanotecnologia e Fotonica per la Medicina di Precisione' – CUP B83B17000010001. L.L.d.M. acknowledges the European Union (ERC-StG project INTERCELLMED, Contract No. 759959) for partial funding. A.B. acknowledges supports from the Match Network through Progetto Pythagoras (CUP:J48C17000250006) and partial (basal) support from INFN and MIUR. Y.S.Z. acknowledges supports by the National Institutes of Health (K99CA201603, R21EB025270) and the New England Anti-Vivisection Society (NEAVS).

## References

1 Bhatia, S.N. and Ingber, D.E. (2014) Microfluidic organs-on-chips. *Nat. Biotechnol.* 32, 760–772
2 Polini, A. *et al.* (2014) Organs-on-a-chip: a new tool for drug discovery. *Expert Opin. Drug Discov.* 9, 335–352
3 Esch, E.W. *et al.* (2015) Organs-on-chips at the frontiers of drug discovery. *Nat. Rev. Drug Discov.* 14, 248–260
4 Low, L.A. and Tagle, D.A. (2018) 'You-on-a-chip' for precision medicine. *Expert Rev. Precis. Med. Drug Dev.* 3, 137–146
5 Yesil-Celiktas, O. *et al.* (2018) Mimicking human pathophysiology in organ-on-chip devices. *Adv. Biosyst.* 2, 1800109 http://dx.doi.org/10.1002/adbi.201800109 Published online August 13, 2018
6 Balijepalli, A. and Sivaramakrishan, V. (2017) Organs-on-chips: research and commercial perspectives. *Drug Discov. Today* 22, 397–403
7 Watson, D.E. *et al.* (2017) Fitting tissue chips and microphysiological systems into the grand scheme of medicine, biology, pharmacology, and toxicology. *Exp. Biol. Med.* 242, 1559–1572
8 Zhang, Y.S. *et al.* (2017) Multisensor-integrated organs-on-chips platform for automated and continual in situ monitoring of organoid behaviors. *Proc. Natl. Acad. Sci. U. S. A.* 114, E2293–E2302
9 Rich, R.R. *et al.* (2019) *Clinical Immunology* (5th edn.), Elsevier
10 Guan, M. *et al.* (2015) Adverse events of monoclonal antibodies used for cancer therapy. *Biomed. Res. Int.* 2015, 428169
11 Baldo, B.A. (2013) Adverse events to monoclonal antibodies used for cancer therapy: Focus on hypersensitivity responses. *Oncoimmunology* 2, e26333
12 Ribas, A. and Wolchok, J.D. (2018) Cancer immunotherapy using checkpoint blockade. *Science* 359, 1350–1355
13 Batlevi, C.L. *et al.* (2016) Novel immunotherapies in lymphoid malignancies. *Nat. Rev. Clin. Oncol.* 13, 25–40
14 June, C.H. *et al.* (2018) CAR T cell immunotherapy for human cancer. *Science* 359, 1361–1365
15 Kroschinsky, F. *et al.* (2017) New drugs, new toxicities: severe side effects of modern targeted and immunotherapy of cancer and their management. *Crit. Care* 21, 89
16 Norelli, M. *et al.* (2018) Monocyte-derived IL-1 and IL-6 are differentially required for cytokine-release syndrome and neurotoxicity due to CAR T cells. *Nat. Med.* 24, 739–748
17 Giavridis, T. *et al.* (2018) CAR T cell-induced cytokine release syndrome is mediated by macrophages and abated by IL-1 blockade. *Nat. Med.* 24, 731–738
18 Brehm, M.A. *et al.* (2016) Humanized mice in translational immunology. In *Translational Immunology* (Tan, S.-L., ed.), pp. 285–326, Elsevier
19 Walsh, N.C. *et al.* (2017) Humanized mouse models of clinical disease. *Annu. Rev. Pathol.* 12, 187–215
20 Murphy, K. *et al.* (2012) *Janeway's Immunobiology* (8th ed.), Garland Science
21 Mercier, F.E. *et al.* (2011) The bone marrow at the crossroads of blood and immunity. *Nat. Rev. Immunol.* 12, 49–60
22 Nombela-Arrieta, C. and Manz, M.G. (2017) Quantification and three-dimensional microanatomical organization of the bone marrow. *Blood Adv.* 1, 407–416
23 Irimia, D. and Ellett, F. (2016) Big insights from small volumes: deciphering complex leukocyte behaviors using microfluidics. *J. Leukocyte Biol.* 100, 291–304
24 Shao, N. and Qin, L. (2018) Biochips-new platforms for cell-based immunological assays. *Small Methods* 2, 1700254
25 Boussommier-Calleja, A. *et al.* (2016) Microfluidics: a new tool for modeling cancer-immune interactions. *Trends Cancer* 2, 6–19
26 Huang, C.P. *et al.* (2009) Engineering microscale cellular niches for three-dimensional multicellular co-cultures. *Lab Chip* 9, 1740–1748
27 Ayuso, J.M. *et al.* (2016) Development and characterization of a microfluidic model of the tumour microenvironment. *Sci. Rep.* 6, 36086
28 Moore, N. *et al.* (2018) A multiplexed microfluidic system for evaluation of dynamics of immune-tumor interactions. *Lab Chip* 18, 1844–1858
29 Pavesi, A. *et al.* (2017) A 3D microfluidic model for preclinical evaluation of TCR-engineered T cells against solid tumors. *JCI Insight* 2, 89762
30 Han, S. *et al.* (2012) A versatile assay for monitoring *in vivo*-like transendothelial migration of neutrophils. *Lab Chip* 12, 3861–3865
31 Ingram, P.N. *et al.* (2018) An accessible organotypic microvessel model using iPSC-derived endothelium. *Adv. Healthc. Mater.* 7, 1700497
32 Park, J.Y. *et al.* (2011) Monitoring the status of T-cell activation in a microfluidic system. *Analyst* 136, 2831–2836
33 Wu, W.-H. *et al.* (2015) A capillary-endothelium-mimetic microfluidic chip for the study of immune responses. *Sens. Actuators B Chem.* 209, 470–477
34 Menon, N.V. *et al.* (2017) Micro-engineered perfusable 3D vasculatures for cardiovascular diseases. *Lab Chip* 17, 2960–2968
35 Cucullo, L. *et al.* (2011) A dynamic in vitro BBB model for the study of immune cell trafficking into the central nervous system. *J. Cereb. Blood Flow Metab.* 31, 767–777
36 Sackmann, E.K. *et al.* (2012) Microfluidic kit-on-a-lid: a versatile platform for neutrophil chemotaxis assays. *Blood* 120, e45–e53
37 Kim, H.J. *et al.* (2016) Contributions of microbiome and mechanical deformation to intestinal bacterial overgrowth and inflammation in a human gut-on-a-chip. *Proc. Natl. Acad. Sci. U. S. A.* 113, E7–E15
38 Shah, P. *et al.* (2016) A microfluidics-based in vitro model of the gastrointestinal human-microbe interface. *Nat. Commun.* 7, 11535
39 Benam, K.H. *et al.* (2016) Matched-comparative modeling of normal and diseased human airway responses using a microengineered breathing lung chip. *Cell Syst.* 3, 456–466
40 Dexter, T.M. (1979) Hemopoiesis in long-term bone marrow cultures. A review. *Acta Haematol.* 62, 299–305
41 Whitlock, C.A. and Witte, O.N. (1982) Long-term culture of B lymphocytes and their precursors from murine bone marrow. *Proc. Natl. Acad. Sci. U. S. A.* 79, 3608–3612
42 Di Maggio, N. *et al.* (2011) Toward modeling the bone marrow niche using scaffold-based 3D culture systems. *Biomaterials* 32, 321–329
43 Kim, J. *et al.* (2015) Organ-on-a-chip: development and clinical prospects toward toxicity assessment with an emphasis on bone marrow. *Drug Saf.* 38, 409–418
44 Torisawa, Y.S. *et al.* (2014) Bone marrow-on-a-chip replicates hematopoietic niche physiology *in vitro*. *Nat. Meth.* 11, 663–669
45 Sieber, S. *et al.* (2018) Bone marrow-on-a-chip: long-term culture of human haematopoietic stem cells in a three-dimensional microfluidic environment. *J. Tissue Eng. Regener. Med.* 12, 479–489
46 Bruce, A. *et al.* (2015) Three-dimensional microfluidic tri-culture model of the bone marrow microenvironment for study of acute lymphoblastic leukemia. *PLoS One* 10, e0140506
47 Braham, M.V.J. *et al.* (2018) Cellular immunotherapy on primary multiple myeloma expanded in a 3D bone marrow niche model. *Oncoimmunology* 7, e1434465
48 Bredenkamp, N. *et al.* (2014) An organized and functional thymus generated from FOXN1-reprogrammed fibroblasts. *Nat. Cell Biol.* 16, 902–908
49 Ross, A.E. *et al.* (2017) Spatially resolved microfluidic stimulation of lymphoid tissue ex vivo. *Analyst* 142, 649–659
50 Purwada, A. *et al.* (2015) Ex vivo engineered immune organoids for controlled germinal center reactions. *Biomaterials* 63, 24–34
51 Purwada, A. and Singh, A. (2017) Immuno-engineered organoids for regulating the kinetics of B-cell development and antibody production. *Nat. Protoc.* 12, 168–182
52 Purwada, A. *et al.* (2017) Modular immune organoids with integrin ligand specificity differentially regulate ex vivo B cell activation. *ACS Biomater. Sci. Eng.* 3, 214–225
53 Moura Rosa, P. *et al.* (2016) The intercell dynamics of T cells and dendritic cells in a lymph node-on-a-chip flow device. *Lab Chip* 16, 3728–3740
54 Giese, C. *et al.* (2010) Immunological substance testing on human lymphatic micro-organoids *in vitro*. *J. Biotechnol.* 148, 38–45
55 Seifert, M. *et al.* (2012) Crosstalk between immune cells and mesenchymal stromal cells in a 3D bioreactor system. *Int. J. Artif. Organs* 35, 986–995
56 Radke, L. *et al.* (2017) *In vitro* evaluation of glycoengineered RSV-F in the human artificial lymph node reactor. *Bioengineering* 4, E70 57 Mannino, R.G. et al. (2017) 3D microvascular model recapitulates the diffuse large B-cell lymphoma tumor microenvironment in vitro. Lab Chip 17, 407–414
58 Apoorva, F. et al. (2018) How biophysical forces regulate human B cell lymphomas. Cell Rep. 23, 499–511
59 Kang, J.H. et al. (2014) An extracorporeal blood-cleansing device for sepsis therapy. Nat. Med. 20, 1211–1216
60 Rigat-Brugarolas, L.G. et al. (2014) A functional microengineered model of the human splenon-on-a-chip. Lab Chip 14, 1715–1724
61 Picot, J. et al. (2015) A biomimetic microfluidic chip to study the circulation and mechanical retention of red blood cells in the spleen. Am. J. Hematol. 90, 339–345
62 WaxDesign Corporation (2018) MIMIC® Technology. WaxDesign Corporation
63 Agliari, E. et al. (2014) Cancer-driven dynamics of immune cells in a microfluidic environment. Sci. Rep. 4, 6639
64 Biselli, E. et al. (2017) Organs on chip approach: a tool to evaluate cancer–immune cells interactions. Sci Rep. 7, 12737
65 Agliari, E. et al. (2015) Anergy in self-directed B lymphocytes: a statistical mechanics perspective. J. Theor. Biol. 375, 21–31
66 Burroughs, N.J. et al. (2006) Regulatory T cell adjustment of quorum growth thresholds and the control of local immune responses. J. Theor. Biol. 241, 134–141
67 Busse, D. et al. (2010) Competing feedback loops shape IL-2 signaling between helper and regulatory T lymphocytes in cellular microenvironments. Proc. Natl. Acad. Sci. U. S. A. 107, 3058–3063
68 Butler, T.C. et al. (2013) Quorum sensing allows T cells to discriminate between self and nonself. Proc. Natl. Acad. Sci. U. S. A. 110, 11833–11838
69 Zikherman, J. and Au-Yeung, B. (2015) The role of T cell receptor signaling thresholds in guiding T cell fate decisions. Curr. Opin. Immunol. 33, 43–48
70 Mora, T. et al. (2010) Maximum entropy models for antibody diversity. Proc. Natl. Acad. Sci. U. S. A. 107, 5405–5410
71 Agliari, E. et al. (2012) Multitasking associative networks. Phys. Rev. Lett. 109, 268101
72 Agliari, E. et al. (2013) Immune networks: multitasking capabilities near saturation. J. Phys. A 46, 415003
73 Boneschansker, L. et al. (2018) Convergent and divergent migratory patterns of human neutrophils inside microfluidic mazes. Sci. Rep. 8, 1887
74 Reategui, E. et al. (2017) Microscale arrays for the profiling of start and stop signals coordinating human-neutrophil swarming. Nat. Biomed. Eng. 1, 0094
75 Giese, C. and Marx, U. (2014) Human immunity in vitro – solving immunogenicity and more. Adv. Drug Deliv. Rev. 69–70, 103–122
76 Tononi, G. et al. (1999) Measures of degeneracy and redundancy in biological networks. Proc. Natl. Acad. Sci. U. S. A. 96, 3257–3262
77 Crumiller, M. et al. (2011) Estimating the amount of information conveyed by a population of neurons. Front. Neurosci. 5, 90
78 Singh, A. (2017) Biomaterials innovation for next generation ex vivo immune tissue engineering. Biomaterials 130, 104–110